\documentclass{LMCS}

\def\dOi{10(1:9)2014}
\lmcsheading%
{\dOi}
{1--23}
{}
{}
{Jan.~18, 2013}
{Feb.~12, 2014}
{}


\ACMCCS{[{\bf Theory of computation}]: Models of
  computation---Interactive computation; Logic---Constructive mathematics}

\makeatletter
\DeclareRobustCommand*\cal{\@fontswitch\relax\mathcal}
\makeatother

\usepackage{hyperref}
\usepackage{amsfonts}
\usepackage{amsmath}
\usepackage{amssymb}
\usepackage{amsbsy}

\theoremstyle{plain}
\theoremstyle{plain}\newtheorem{lemma}[thm]{Lemma}
\theoremstyle{plain}\newtheorem{definition}[thm]{Definition}
\theoremstyle{plain}\newtheorem{theorem}[thm]{Theorem}

\newcounter{esemconta}
\newcommand{\iniese}
                    {
                                \refstepcounter{esemconta}
                                \medskip
    \noindent {\bf Example \theesemconta}
                        }

\newcommand{\arrow}{\rightarrow}

\newcommand{\comp}{\circ}  % composition
\newcommand{\Id}{\mbox{\rm id}}

\newcommand{\Set}[1]{\{#1\}}
\newcommand{\Pair}[2]{\langle #1, #2 \rangle}
\newcommand{\Open}{{\cal O}}

\newcommand{\Iff}{\Leftrightarrow}
\newcommand{\Then}{\Rightarrow}

\newcommand{\QED}{\qed}

\newcommand{\Atom}          { { \mathbb{A} } }
\newcommand{\State}         { { \mathbb{S}_{\mbox{\footnotesize \it fin}} } }
\newcommand{\Bool}          {2}
\newcommand{\True}	        {\mbox{\sf T}}
\newcommand{\False}			{\mbox{\sf F}}
\newcommand{\trueVal}	    { {\footnotesize \mbox{\sf true }} }
\newcommand{\falseVal}	    { {\footnotesize \mbox{\sf false}} }

\newcommand{\Nat}           { { \mathbb{N} } }
\newcommand{\InfState}			{ { \mathbb{S} } }

\newcommand{\compat}				{\sim}
\newcommand{\StateTop}			{\Omega(\InfState)}
\newcommand{\Compatible}[2]	{#1 \# #2}
\newcommand{\NotCompatible}[2]	{\neg \, #1 \# #2}

\newcommand{\OpenSet}				{{\cal O}}

\newcommand{\FinPow}			{{\cal P}_{\mbox{\footnotesize \it fin}}}

\newcommand{\Compl}[1]			{\overline{#1}}
\newcommand{\UpSet}[1]			{#1\!\uparrow}
\newcommand{\Compact}[1]		{K(#1)}
\newcommand{\New}						{{\sf new}}
\newcommand{\Level}					{{\sf lev}}
\newcommand{\Ord}						{\mbox{\it Ord}}
\newcommand{\SubOrd}				{\mbox{\footnotesize \it Ord}}

\newcommand{\restr}					{\upharpoonright}
\newcommand{\Operator}			{\mbox{\it Op}}
\newcommand{\Realizer}			{{\cal R}}
\newcommand{\Truth}					{{\sf tr}}

\newcommand{\realz}[1]			{\widehat{#1}}

\newcommand{\Question}           { { \mathbb{Q} } }
\newcommand{\query}              { {\tt q}}
\newcommand{\Sound}              { {\tt S} }
\newcommand{\Complete}           { {\tt C} }
\newcommand{\Model}              { {\tt M} }

\newcommand{\atoms}              {{answers} }

\newcommand{\LocMin}[1]{#1\mbox{\rm -min}}
\newcommand{\real}{\vdash}

\newcommand{\hide}[1]{#1}

\begin{document}

\title[Knowledge Spaces]{Knowledge Spaces and the Completeness of Learning Strategies}

\author[S.~Berardi]{Stefano Berardi\rsuper a}	%required
\address{{\lsuper{a,b}}Dipartimento di Informatica, Universit\`a di Torino
  c.so Svizzera 185 Torino, Italy}	%required
\email{\{stefano,deligu\}@di.unito.it}  %optional
\thanks{{\lsuper{a,b}}This work was partially supported by PRIN project n. 2008H49TEH}	%optional

\author[U.~de'Liguoro]{Ugo de'Liguoro\rsuper b}	%optional
\address{\vspace{18 pt}}	%optional
%\email{deligu@di.unito.it}  %optional
%\thanks{This work was partially supported by PRIN project n. 2008H49TEH.}	%optional

%% mandatory lists of keywords and classifications:

\maketitle

\begin{abstract}
\noindent
We propose a theory of learning aimed to formalize some ideas underlying Coquand's game semantics and Krivine's realizability of classical logic. We introduce a notion of knowledge state together with a new topology, capturing finite positive and negative information that guides a learning strategy. We use a leading example to illustrate how non-constructive proofs lead to continuous and effective learning strategies over knowledge spaces, and prove that our learning semantics is sound and complete w.r.t. classical truth, as it is the case for Coquand's and Krivine's approaches.
 \end{abstract}

\section*{Introduction}

Several methods have been proposed to give a recursive interpretation of non-recursive constructions of mathematical objects,
whose existence and properties are classically provable. A non-exhaustive list includes the continuation-based approach initiated by Griffin \cite{Gri90}, the game theoretic semantics of classical arithmetic by Coquand  \cite{Coq95} and Krivine's realizability of classical logic \cite{KrivineRealizability}.

As observed by Coquand, there is a common informal
idea underlying the different approaches, which is learning. With respect to the dialogic approach, learning consists into interpreting the strategy for the defender of a statement against the refuter by a strategy guiding the interaction between a learning agent and the ``world'', representing what can be experienced by direct computation.

Under the influence of Gold's ideas \cite{Gold65,Gold67} and of Hayashi's Limit Computable
Mathematics \cite{Hay06} we have proposed a formal theory of ``learning'' and ``well founded limits'' in \cite{Berardi-deLiguoro:IC09}. In the theory the goal of the learning process is to find an evidence, or a witness as it is usually called, of the truth of some given sentence, which is the ``problem'' that the learning strategy solves. Such an evidence is always tentative, since certainty can be attained only in the ideal limit. The task of the learning strategy is to tell how to react to the discovery that the current guess is actually wrong, and this is done on the basis of the knowledge collected in the learning process, which includes all the ``counterexamples'' that have been seen up to the time.

Here we propose the same idea, but in the different perspective of topological spaces and continuous maps.
We assume having an ideal object being the result of some non-effective mental
construction and satisfying some decidable property. We say that this
construction may be learned w.r.t. the property if we can find a
``finite approximation'' of the construction which still satisfies the property.
In particular given a classical proof of an existential statement, we see the computational content of the proof as the activity of guessing more and more about the object (individual) the statement is about, without ever obtaining a full information; in a discrete setting such as natural numbers, the approximation is actually a different object, which we think as ``close'' to the ideal one, the ``limit'', only with respect to some given property which both satisfy.

When reasoning about ideal objects, we deal with descriptions rather than with the objects themselves. While learning of an ideal object we have step by step certain amounts of knowledge, which consist of pieces of evidence (e.g. decidable statements): therefore we topologize {\em states of knowledge} to express the idea that a continuous strategy only depends on finite positive and negative information to yield finite approximations of the ideal limit. We call {\em interactive realizer} any continuous function of states of knowledge that roughly tells which are the further guesses to improve the given knowledge, and how to react to the discovery of negative evidences (e.g. counterexamples to certain assumptions).
We conjecture that interactive realizers can be seen as winning strategies in the sense of Coquand and as the semantics of
lambda terms with continuations in Griffin's work and of terms in a call-by-value version of Krivine's classical realizability interpretation;
however this far from being evident and it deserves further investigation.

We call a {\em model} any perfect (usually infinite) knowledge state.
The main result of this paper is that any object that can be ideally learned in a model can be effectively learned in a finite state of knowledge approximating the model, and this state is found by means of a realizer. Since models represent classical truth, the completeness theorem can be read as stating that the learning semantics for classical proofs is complete w.r.t. classical truth, namely Tarskian truth, as it is the case for Coquand's dialogic semantics of proofs and for Krivine's classical realizability, and that the learning process is indeed effective.

\medskip The paper is organized as follows.
In \S \ref{sec:learning} we introduce a motivating example, which is used throughout the paper.
In \S \ref{sec:state_topology} we define the structures of states of knowledge and their topology. In
\S \ref{sec:leyred} the concept of relative truth is introduced to define sound, complete and model knowledge states. In
\S \ref{sec:realizer} we define interactive realizability and prove the completeness theorem. Finally in \S \ref{sec:algorithm}
we show, although not in full generality, how the existence of a computable realizer over a computable space of knowledge
gives rise to an algorithm to effectively learn approximated solutions.

\subsection*{Related works}

The suggestion by Coquand that the dialogic interpretation of classical proofs could be seen as learning of some abstract entities can be found in \cite{Coq91}, a preliminary version of \cite{Coq95}.
The idea has been illustrated by means of a suggestive analysis of the proof of Sylvester's Conjecture by von Plato in \cite{Plato05}. Beside Krivine's  \cite{KrivineRealizability}, Miquel's work in \cite{Miquel11b} illustrates in detail the behaviour of classical realizability of existential statements (also in comparison to Friedman's method), and we strongly believe that the construction is a learning process in the sense of the present paper.

Learning in the limit of undecidable properties and ideal entities comes from Gold's work \cite{Gold65, Gold67}, and has been recently formulated for classical logic by Hayashi e.g. in \cite{Hay06}. Hayashi defined realizers as limits over a linearly ordered knowledge set and in the case of monotonic learning: the first model with realistic assumptions over of the knowledge states, again in the case of monotonic learning, is in \cite{Ber05}.
We have investigated more the concept of learning in the limit incorporating Coquand's ideas in \cite{Berardi-deLiguoro:IC09}, although in a combinatory rather than topological perspective. We have further elaborated the concept of learnable in the limit in \cite{Berardi-deLiguoro:TOCL-11}, where a ``solution'' in terms of the following \S  \ref{sec:realizer} is called ``individual'', since we identify the ideal limit with the map reading the learned values out states of knowledge. The formal definition of the state topology, however, is new with the present work, as well as the treatment of the general, non-monotonic case. Also the concept of ``interactive realizability'' has been introduced in   \cite{Berardi-deLiguoro:TOCL-11}, but in the simpler case of monotonic learning. Essentially the same construction as in  \cite{Berardi-deLiguoro:TOCL-11} is used in \cite{AschBerLMCS2010} to define a realizability interpretation of {\bf HA} plus excluded middle restricted to $\Sigma^0_1$-formulas. It turns out that interactive realizability is a generalization of Kleene's realizability, and this motivates the terminology. A geometrical application of interactive realizability, akin to Von Plato's work on Sylvester's Conjecture, can be found in Birolo's ph.d. thesis \cite{Bir13}. Recently Berardi and Steila analyzed Erd\"os' proof of Ramsey Theorem \cite{BerSteila13}, where it is shown that there exists an interpretation of such a proof using a three-level non-monotonic learning structure.

\section{Solving problems by learning}\label{sec:learning}
To illustrate the idea of learning strategies, either monotonic or non-monotonic, we propose an example suggested by Coquand and developed by
Fridlender in \cite{Fridlender96}. Let $f_1,f_2$ be total functions over $\Nat$. Fix some integer $k > 0$ and consider the statements: ``there is an increasing sequence of $k$ integers which is weakly increasing w.r.t. $f_1$'' and ``there is an increasing sequence of $k$ integers which is weakly increasing w.r.t. $f_1$ and $f_2$''. Formally:
\begin{eqnarray}
\exists x_1 \ldots \exists x_k. \; x_1 < \cdots < x_k ~\wedge~ f_1(x_1) \leq \cdots \leq f_1(x_k) \label{eq:monotonic}\\ [1mm]
\exists x_1 \ldots \exists x_k. \; x_1 < \cdots < x_k ~\wedge~ f_1(x_1) \leq \cdots \leq f_1(x_k) ~\wedge~  f_2(x_1) \leq \cdots \leq f_2(x_k)
\label{eq:non-monotonic}
\end{eqnarray}

We look at these statements as the {\em problems} of finding a $k$-tuple $n_1 < \cdots < n_k$ witnessing their truth.
We begin by observing that these statements can be proved classically as follows. For any $f:\Nat\arrow\Nat$ and $A\subseteq\Nat$ we say that ``$n$ is a local minimum of $f$ w.r.t. $A$'' (shortly $n$ is an $f,A$-minimum) if $n$ is the minimum of $f$ in $A \cap [n,\infty[$. Formally:
\[\LocMin{f,A}(n) \Iff \forall y\in A. \ n < y \Then f(n) \leq f(y).\]
Observe that the predicate $\LocMin{f,A}(n)$ is undecidable in general, even if $f$ is recursive and $A$ decidable. The statement $\neg \LocMin{f,A}(n)$ is classically equivalent to $\exists y\in A. \ n < y \wedge f(n) > f(y)$. For any $f:\Nat\arrow\Nat$ and infinite $A \subseteq \Nat$, we denote by
\[A_f = \Set{a \in A \mid \LocMin{f,A}(a)} \subseteq A\]
the set of all $f,A$-minima. We now study the set $A_f$.

\begin{lemma}\label{lem:f-min}
For any $f:\Nat\arrow\Nat$ and infinite $A \subseteq \Nat$, the set
$A_f = \Set{a \in A \mid \LocMin{f,A}(a)}$
of $f,A$-minima is infinite, and $f$ is monotonic over $A_f$.
\end{lemma}

\proof
Toward a contradiction suppose that $A_f$ is finite, possibly empty. Then there exists $a_0 = \min (A\setminus A_f)$, as $A$ is infinite. By definition of $a_0$ for all $a \in A$ with $a \ge a_0$ we have $\neg \LocMin{f,A}(a)$, that is: there is some $a' > a$, $a' \in A$ such that $f(a) > f(a')$. If we choose $a = a_0$ we deduce that there exists some $a_1 \in A$  such that $a_0 < a_1$ and  $f(a_0) > f(a_1)$, and so on. By iterating the reasoning we get an infinite sequence $a_0 < a_1 < a_2 < \cdots$ such that $f(a_i) > f(a_{i+1})$ for all $i \in \Nat$, which is on turn an infinite descending chain in $\Nat$, that is an absurdity.
\QED

\begin{theorem}\label{prop:f-minima} Both statements (\ref{eq:monotonic}) and (\ref{eq:non-monotonic}) are classically provable.
\end{theorem}

\proof In Lemma \ref{lem:f-min} take $A = \Nat$: then $f_1$ is monotonic over the infinite set $A_{f_1}$. If we take the first $k$ elements of $A_{f_1}$ we have an increasing sequence of $k$ integers whose values are weakly increasing under $f_1$, establishing  (\ref{eq:monotonic}).

To prove (\ref{eq:non-monotonic}) we use use again Lemma \ref{lem:f-min}.2 taking  $A = \Nat_{f_1}$, which we know to be infinite by the same lemma; then $f_1, f_2$ are both monotonic  over the infinite set $A_{f_2} = (\Nat_{f_1})_{f_2}$.
\QED

\medskip
 The proof of Lemma \ref{lem:f-min} and its use in the proof of Theorem \ref{prop:f-minima} are non-constructive, as they rely on the ``computation'' of the minimum of $f$ in $A$
for certain $f$ and $A$. In order to compute the minimum of $f$ we need, in general, to know infinitely many values of $f$.
However, this proof may be interpreted by a computation
as soon as we only require finitely many $n_1, \ldots , n_k$ that satisfy (\ref{eq:monotonic}) or (\ref{eq:non-monotonic}). This
is because $n_1, \ldots , n_k$ can be found using a finite knowledge about $f$.

Indeed the proof of Lemma \ref{lem:f-min} can be turned into an effective strategy to learn a solution to problem (\ref{eq:monotonic}), assuming that $f_1$ is recursive. The basic remark is that we do not actually need to know the infinitely many elements of $\Nat_{f_1}$, nor we have to produce some $n$ which belongs to $\Nat_{f_1}$ {\em beyond any doubt}. We can approximate the infinite set $\Nat_{f_1}$ by some finite set $B$, whose elements are not necessarily in $\Nat_{f_1}$, rather they are just $f_1,\Nat$-minima  {\em as far as we know}. $B$ is a kind of hypothesis about $\Nat_{f_1}$.

More precisely, we will find a set $B$ such that $\LocMin{f_1,B}(n)$ for all $n \in B$. This property is trivially true for any singleton set, say $\Set{0}$. In the general case, if $B$ has $k$ elements or more, then by definition of $\LocMin{f_1,B}$ the set $B$ is a solution to (\ref{eq:monotonic}). Otherwise we take any $m >  \max(B)$ and try adding $m$ to $B$. Since we cannot decide whether any $n$ is a local minimum of $f_1$, we are not allowed to increase $B$ to $B\cup\Set{m}$, because it could be the case that
$f_1(n) > f_1(m)$ for some $n \in B$. Rather we update $B$, by removing all $n \in B$ such that $f_1(n) > f_1(m)$:
\[ B' = \Set{n \in B \mid f_1(n) \leq f_1(m)} \cup \Set{m} ,\]
The new set $B'$ includes $m$ and satisfies the invariant property of containing only ${f_1,B'}$-minima.
The cardinality of $B'$ is not necessarily greater than that of $B$, so that we need an argument to conclude that, starting from the singleton $\Set{0}$ and iterating the step from $B$ to $B'$, the learning agent will eventually reach a $k$-element set with the required property.

The termination argument in this case works as follows. Although the sequence of sets is not increasing w.r.t. inclusion, the knowledge that some elements are {\em not}  local minima of $f_1$ grows monotonically, since more and more pairs $n, m$ are found such that $f_1(n) > f_1(m)$. From this remark, one can prove by a fixed-point argument (over a suitable topology) that the growth of knowledge eventually ends, which implies that a set $B$ with $k$ elements will be found after finitely many steps. In this case we speak of {\em monotonic learning}.

In order to include an example of non-monotonic learning, we assume that both $f_1$ and $f_2$ are recursive, and we outline an effective computation approximating  an initial segment of $(\Nat_{f_1})_{f_2}$, solving problem (\ref{eq:non-monotonic}).
The informal interpretation we include here will be formalized in the following sections using the notion of layered valuation.

As it happens in the classical proof of Theorem
\ref{prop:f-minima}, we iterate the same method used for problem (\ref{eq:monotonic}), and build a $C \subseteq B$ of $f_2,C$-minima, where $B$ is the current approximation of the infinite set $\Nat_{f_1}$. In doing so the learning agent assumes that $B$ is a subset of $\Nat_{f_1}$ (though she cannot be certain of this), and that all elements of $C$ are $f_2$-minima w.r.t. $\Nat_{f_1}$ (again an uncertain belief). At each step, the learner takes some $m \in B$, such that $m > \max(C)$, and manages to add $m$ to $C$, possibly by removing some of its elements, by computing:
\[ C' = \Set{p \in C  \mid f_2(p) \leq f_2(m)} \cup \Set{m} \subseteq B .\]
This is only possible when such an $m$ exists in $B$: if not the algorithm generating $B$ has to be resumed to get a larger set containing an element greater than $ \max(C)$. But since $B$ does not grow monotonically, elements of $C$ will be dropped while computing $C'$ also because they are no longer in $B$. This makes the convergence proof much harder.
Indeed the knowledge accumulated while building $B$ takes the simple form of (sets of statements) $f_1(n) > f_1(m)$, and it grows monotonically; on the contrary the ``knowledge'' gathered while computing $C$ consists of more complex statements of the form: $m \in B \wedge f_2(n) > f_2(m)$, with $B$ changing non-monotonically during the computation of $C$. This knowledge is the conjunction of the hypothesis (uncertain beliefs) $m \in B$ and the fact $f_2(n) > f_2(m)$ (a decidable statement). This second layer of knowledge, mixing hypothesis and facts, does not grow monotonically, because any hypothesis $m \in B$ may turn out to be false: therefore it is unsafe, yet it guides the construction of $C$.
In this case we speak of {\em non-monotonic learning}.
Non-monotonic learning is the most general form of learning.

\section{States of knowledge and their topology}
\label{sec:state_topology}

There are three kinds of entities in learning: questions, answers and states of knowledge. The main concern are states of knowledge, which on turn are certain sets of answers. Answers are viewed as atomic objects although in the examples they are logical formulas,
since their internal structure is immaterial from the point of view of the topology of states of knowledge we are introducing here, while their complexity is abstractly represented by levels in the next section.
Questions instead are represented indirectly by equivalence classes of answers,
each to be thought of as the set of alternative, incompatible choices for an answer to the same question.

\begin{definition} [Knowledge Structure and State of Knowledge]\label{def:knowledgeState}
A {\em knowledge structure} $(\Atom, \compat)$ consists of a non-empty countable set $\Atom$ of {\em answers}
and an equivalence relation $\compat \;\subseteq \Atom \times \Atom$. As a topological space, $\Atom$ is equipped with the discrete topology.

The set $\Question = \Atom/_{\compat}$ of equivalence classes $[x]$ w.r.t. $\compat$ is the set of {\em questions}, and it is equipped with the discrete topology.

A subset $X \subseteq \Atom$ is a {\em state of knowledge} if for all $x\in \Atom$ the set $X\cap [x]$ is either empty or a singleton. We denote by $\InfState$ the set of knowledge states and by
$\State$ the subset of finite elements of $\InfState$.
\end{definition}

If $x \compat y$ then $x,y$ are two answers to the same question. The equivalence class $[x]$ abstractly represents the question answered by $x$.
Two \atoms $x, y \in \Atom$ are {\em compatible}, written $\Compatible{x}{y}$, if they are not different answers to the same question, namely if:
\[x  \compat y \Then x = y.\]
%\[\Compatible{x}{y} \; \Iff \; x = y \OR x \not \compat y.\]
Now it turns out that $\InfState$ is the coherence space whose web is the graph $(\Atom, \#)$ (see \cite{girard:1989a}
\S 8.2). Indeed
we observe that $\InfState$ is a poset by subset inclusion, and it is downward closed. It follows that
$(\InfState, \cap, \subseteq)$ is an inf-semilattice with bottom $\emptyset$.
$\InfState$ is closed under arbitrary
but non-empty inf, as the empty inf, namely the whole set $\Atom$, is not consistent in general.
Indeed $\InfState$
is not closed under union, unless the compatibility relation is the identity.
We say that $X$ and $Y$ are compatible w.r.t. inclusion,
if $X \subseteq Z \supseteq Y$ for some
$Z\in \InfState$. Clearly
the union of a family ${\cal U}\subseteq \InfState$ belongs to $\InfState$ if and only
if all elements of ${\cal U}$ are pairwise compatible sets.
By this $\InfState$ is closed under directed and in general bounded sups,
so it is a domain which in fact has compacts $\Compact{\InfState} = \State$ and  is algebraic.

\iniese{Let us reconsider the example  in \S \ref{sec:learning}. A knowledge structure $(\Atom_0,\compat_0)$
for learning a solution to (\ref{eq:monotonic}) can be defined by taking $\Atom_0 = \Set{(n,m)\in \Nat\times\Nat \mid n < m}$, where we interpret a pair $(n,m)$ as the statement: ``$m$ is a counterexample to
$\LocMin{f_1,\Nat}(n)$'', and more precisely as the formula:
\[ n < m \wedge f_1(n) > f_1(m). \]
If we think of $m$ as the answer to the question about $n$, we obtain the definition of the relation $(n,m) \compat_0 (n',m')$ by $n = n'$.

A knowledge structure $(\Atom_1,\compat_1)$
for learning a solution to (\ref{eq:non-monotonic})
can be defined by taking $\Atom_1 = \Atom_0$ and $\compat_1 \,=\, \compat_2$. However
$(\Atom_1,\compat_1)$ has not the same intended meaning as $(\Atom_0,\compat_0)$; indeed an answer $(n,m)\in\Atom_1$ is interpreted by the statement:
\[n < m \wedge f_2(n) > f_2(m) \wedge n,m \in \Nat_{f_1}.\]

Finally we set $\Atom_2 = \Atom_0 \uplus \Atom_1 = \Set{(i,n,m) \mid i \in \Set{0,1} \And (n,m)\in \Atom_i}$, namely
the disjoint union of $\Atom_0$ and $\Atom_1$ and we define $(i,n,m) \compat_2 (j,n',m')$ if and only if $i = j$ and $n = n'$ (that is
$(n,m) \compat_i (n',m')$).

Let $\InfState_2$ be the knowledge space associated to $\Atom_2$ and $X\in \InfState_2$  be any state of knowledge. Then we interpret $(0,n,m) \in X$ by ``the agent knows at $X$ that $n<m$ and $f_1(n)>f_1(m)$'', hence that $n$ is not in $\Nat_{f_1}$. We interpret: for all $p \in \Nat$, $(0,n,p) \not\in X$ by ``the agent knows at $X$ of no $p$ such that $n<p$ and $f_1(n)>f_1(p)$'', hence she believes that $n$ is in $\Nat_{f_1}$. We write $(\Nat_{f_1})^X = \Set{n \in \Nat \mid \forall p \in \Nat. (0,n,p) \not \in X }$ for the set of $n$ which the agent believes to be in $\Nat_{f_1}$ at $X$. In the same way we write $( (\Nat_{f_1})_{f_2} )^X = \Set{n \in (\Nat_{f_1})^X \mid \forall p \in \Nat. (1,n,p) \not \in X}$ for the set of $n$ which the agent believes to be in  $(\Nat_{f_1})_{f_2}$ at $X$.
We will see in the following sections that $(\Atom_2,\compat_2)$ is a knowledge structure apt
to learn  (\ref{eq:non-monotonic}).

}\label{ex:knowledge-structure}

\medskip
In a state of knowledge $X$ two distinct answers $x,y \in X$ must be compatible, that is $x \not\compat y$.
We consider two answers to the same question as incompatible because we see them as the outcome of a computation process, that returns at most one answer to each question. In this perspective a question can be figured as a memory cell, whose value is either undefined or just one, and that can change in time.
A knowledge state is an abstraction of the state of the whole memory, hence recording only compatible answers.

The ``state of knowledge'' of a finite agent should be finite; for the sake of the theory we also consider infinite states of knowledge, which are naturally approximated by finite ones in a sense to be made precise by a topology. Beside these conceptual reasons, the topological treatment that follows will be essential in the proofs of Theorems
\ref{thr:soundFiniteZero} and \ref{thr:sound-complete}, as well as in the proof of Theorem \ref{thr:algoSoundTerm} establishing convergence of
the algorithm in section \ref{sec:algorithm}.

Let us define a query map $\query: \InfState \times \Question \rightarrow \FinPow(\Atom)$, where  $\FinPow(\Atom)$ is the set of finite subsets of $\Atom$, by $\query(X,[x])=X \cap [x]$. Then $\query(X,[x])$ is either a singleton $\Set{y}$, meaning that, at $X$, $y$ is the answer to the question $[x]$, or the empty set,
meaning that the agent knows at $X$ of no answer to $[x]$, and she assumes that there is none.
Take the discrete topology over $\FinPow(\Atom)$; we then consider the smallest topology over $\InfState$ making $\query$ continuous.

\begin{definition}[State Topology]\label{def:state_topology}
The {\em state topology} $(\InfState, \StateTop)$ is generated by the subbasics $A_x, B_x$, with $x\in\Atom$:
\[\begin{array}{lll}
A_x & = & \Set{X \in \InfState \mid x \in X} =  \Set{X \in \InfState \mid X \cap [x] = \Set{x}},\vspace{2mm}\\
B_x & = & \Set{X \in \InfState \mid X \cap [x] = \emptyset}.
\end{array}\]
\end{definition}

\noindent
Let $X,Y,Z$ range over $\InfState$, and $s,t$ over $\State = \InfState \cap \FinPow(\Atom)$. By definition, a basic open of $\StateTop$ has the shape:
\[\OpenSet_{U,V} = \bigcap_{x\in U} A_x \cap \bigcap_{y\in V} B_y,\]
for finite $U,V \subseteq \Atom$. If $\NotCompatible{x}{y}$, that is $x \compat y$ and $x \neq y$, then $A_x \cap A_y = \emptyset$, because if $x, y \in X$ then $X$ would be inconsistent, so that $\emptyset$ is a basic open. On the other hand if $x \compat y$ then $B_x = B_y$. Therefore without loss of generality we assume $U,V$ to be consistent, and that all basic opens of $\StateTop$ are of the form $\OpenSet_{U,V}=\OpenSet_{s,t}$, for some $s,t\in\State$.
Summing up, assume that $s = \Set{x_1,\ldots,x_n}$ and $t = \Set{y_1,\ldots,y_m}$. Then $X \in \OpenSet_{s,t}$ means that at $X$ the agent knows the answers $x_1,\ldots,x_n$ to the questions $[x_1], \ldots, [x_n]$ (a finite positive information), while she knows of no answer to the questions $[y_1],\ldots,[y_m]$, and assumes that there is none (a finite negative information).

The same thing can be looked at by observing that the continuous maps $f: \InfState \rightarrow I$, with $I$ a discrete space, are those ``asking finitely many questions to their argument''. By this we mean: $f$ is continuous if and only if for all $X \in \InfState$ there are finitely many questions $[x_1], \ldots, [x_n] \in \Question$ such that for all $Y \in \InfState$, if $\query(X,[x_i]) = \query(Y,[x_i])$ for $i=1,\ldots, n$ then $f(X) = f(Y)$.

The state topology is distinct from, yet strictly related to, several well-known topologies.
$\StateTop$ is discrete if and only if $\Question$ is finite (there are finitely many equivalence classes).
$\StateTop$ is homeomorphic to the product space $\Pi_{[x] \in \Question} ([x]\uplus 1)$ of the discrete topologies over $[x]\uplus 1$, where $1$ is any singleton (representing the ``undefined'' answer to the question $[x]$) and $\uplus$ is disjoint union.  Every state topology is homeomorphic to some subspace of the Baire topology over $\Nat^{\Nat}$. The state topology is totally disconnected and Hausdorff: it is compact (namely it is a Stone space)
if and only if all equivalence classes are finite.
If all equivalence classes in $\Question$ are singletons (that is, if $\sim$ is the equality relation on $\Atom$) and $\Question$ is infinite then $\StateTop$ is homeomorphic to the Cantor space $2^{\Nat}$. If all equivalence classes in $\Question$ are infinite and $\Question$ is infinite, then $\StateTop$ is homeomorphic to the whole Baire space.

A clopen is an open and closed set; hence clopens are closed under complement, finite unions and intersections. There are significative examples of clopen sets in $\InfState$.

\enlargethispage{\baselineskip}
\begin{lemma} [Subbasic opens of State Topology are clopen] \label{lem:complBxIsOpen}
Assume that $x \in \Atom$, $f:\InfState \rightarrow I$ is continuous, $I$ is a discrete space and $J \subseteq I$.
Then:
\begin{enumerate}
\item\label{lem:complBxIsOpen-i}
$B_x$ is clopen
\item\label{lem:complBxIsOpen-ii}
$A_x$ is clopen
\item\label{lem:complBxIsOpen-iii}
$f^{-1}(J)$ is clopen.
\end{enumerate}
\end{lemma}

\hide{
\proof\hfill
\begin{enumerate}
\item
It is enough to prove that $\InfState\setminus B_x$ is equal to the union  $\bigcup\Set{A_y\mid y \in [x]}$
of open sets.
Let $X\in\InfState$ and $x\in\Atom$, then:
\[\begin{array}{llll}
X\not\in B_x & \Iff & X \cap [x] \neq \emptyset \\
& \Iff & \exists y\in X. \ x \compat y \\
& \Iff & \exists y. \ y \in [x] \And X \in A_y
\end{array}\]
\item
$S \setminus A_x$ is some union of open
sets. Indeed we have
\[X \not\in A_x \Iff X \in B_x \cup \bigcup\Set{A_y\mid y \in [x] \And y\neq x}.\]
\item
We have $\InfState \setminus f^{-1}(J) = f^{-1}(I \setminus J)$. Both $f^{-1}(J), f^{-1}(I \setminus J)$ are open because they are the counter-image under some continuous $f$ of $J, (I \setminus J)$, which are open sets because $I$ is discrete.\qed
\end{enumerate}
}

\noindent As a consequence of Lemma \ref{lem:complBxIsOpen}.\ref{lem:complBxIsOpen-i} we have that the predicate $n\in\Nat_{f_1}^X$ is continuous in $X$, since $\Set{X \mid n\in\Nat_{f_1}^X} = B_{(0,n,n+1)}$ which is a clopen set, and indeed $n\not\in\Nat_{f_1}^X$ if and only if $X\in \bigcup_{m>n} A_{(0,n,m)}$, namely the complement of $B_{(0,n,n+1)}$. A similar remark holds for
$n\in(\Nat_{f_1})_{f_2}^X$.

\medskip
It is instructive to compare the state topology to Scott and Lawson topologies over $\InfState$.
The {\em Scott topology} over the cpo $(\InfState, \subseteq)$ is determined by taking all the $A_x$ with $x \in \Atom$ as subbasics. On the other hand any $B_x$ is a Scott-closed set, since its complement $\InfState \setminus B_x$ is equal to the union $\bigcup \Set{A_y \mid y \in [x]}$ of Scott opens. However $B_x$ is not Scott-open because it is not upward closed.

Recall that (see \cite{GHK:03}) the {\em lower topology} over a poset is generated by the complements of principal filters;
the {\em Lawson topology} is the smallest refinement of both
the lower and the Scott topology.
In case of the cpo $(\InfState, \subseteq)$ the Lawson topology is generated by the subbasics:
\[ \Compl{\UpSet{X}} = \Set{Y \in \InfState \mid X \not\subseteq Y} ~~~ \mbox{and} ~~~
\UpSet{s}\;=\Set{Y \in \InfState \mid s \subseteq Y},\]
for $X\in \InfState$ and $s\in\State$, representing the negative and positive information respectively.

The state topology includes the Lawson topology, and in general it is finer than that. The next lemma tells that all Lawson opens are open in the state topology, but if some equivalence class $[x]$ is infinite then some open of the state topology is not open in the Lawson topology. Recall that $\StateTop$ denotes the family of open sets in the state topology.

\begin{lemma}\label{lem:state-Lawson}
~
\begin{enumerate}
\item \label{lem:state-Lawson-i}
All basic opens of Lawson topology are in $\StateTop$.
\item \label{lem:state-Lawson-ii}
For all $x \in \Atom$, $B_x$ is Lawson-open if and only if $[x]$ is finite.
\end{enumerate}
\end{lemma}

\proof\hfill
\begin{enumerate}
\item% (\ref{lem:state-Lawson-i})
Since $\UpSet{X} \; = \bigcap_{x\in X} \UpSet{x}$, writing $\UpSet{x}$ for $\UpSet{\Set{x}}$, we have
$\Compl{\UpSet{X}} = \bigcup_{x\in X} \Compl{\UpSet{x}}$, hence it suffices to prove that
$\Compl{\UpSet{x}}\in \StateTop$, for any $x \in \Atom$; since $\Compl{\UpSet{x}} = \Set{Y\in\InfState\mid x \not \in Y} $, this is established by the equality:
\[
\begin{array}{lll}
\Compl{\UpSet{x}} & = & \Set{Y\in\InfState\mid Y \cap [x] = \emptyset}
	\cup \bigcup_{\NotCompatible{x}{y}}\Set{Z \in \InfState\mid Z \cap [x] = \Set{y}} \\
& = & B_x \cup \bigcup_{\NotCompatible{x}{y}} A_y
\end{array}
\]
where $\NotCompatible{x}{y}$ is equivalent to $x \not = y$ and $[x] = [y]$.

\item%(\ref{lem:state-Lawson-ii})
 Observe that $B_x = \bigcap_{y \in [x]} \Compl{\UpSet{y}}$; therefore if
$[x] = \Set{x_1,\ldots,x_k}$ then $B_x = \Compl{\UpSet{x_1}}\,\cap\, \cdots\, \cap\, \Compl{\UpSet{x_k}}$ is a finite intersection of open sets in the Lower topology, hence it is open w.r.t. the Lawson topology. Vice versa suppose that $[x]$ is infinite. Toward a contradiction let
\[\emptyset \not = \OpenSet = \Compl{\UpSet{X_1}}\cap \cdots \cap \Compl{\UpSet{X_k}} \cap \UpSet{s_1} \cap \cdots \cap \UpSet{s_h}
\;\; \subseteq \;B_x\]
be a basic non empty open of the Lawson topology. If we take any $Y \in \OpenSet$, the by definition we have $\Set{x_1,\cdots,x_k} \cap Y = \emptyset$ for some $x_1\in X_1, \ldots, x_n \in X_n$, and $U = s_1\cup\cdots\cup s_h\subseteq Y$ is finite consistent. By the hypothesis that $[x]$ is infinite there is some
$y \in [x] \setminus \Set{x_1,\cdots,x_k} $. By the assumption that $U \subseteq Y \in \OpenSet \subseteq B_x$,
we have that $U \cap [x] \subseteq Y \cap [x] = \emptyset$, hence $V = U \cup \Set{y} \in \State$, namely $V$ is consistent. Now:
\[V \cap \Set{x_1,\cdots,x_k} = (U \cup \Set{y}) \cap \Set{x_1,\cdots,x_k} \subseteq (Y \cup \Set{y}) \cap \Set{x_1,\cdots,x_k} = \emptyset\]
From $U \subseteq V$ and $V \cap \Set{x_1,\cdots,x_k} = \emptyset$ we
deduce that $V \in \OpenSet$, but $V\not\in B_x$ since $y \in V \cap
[x]$: a contradiction.\qed
\end{enumerate}

\noindent As an immediate consequence of Lemma \ref{lem:state-Lawson} we have the following.

\begin{theorem}[State versus Lawson Topology]\label{prop:state-Lawson}
The state topology $\StateTop$ refines the Lawson topology over the cpo $(\InfState, \subseteq)$, and they
coincide if and only if $[x]$ is finite for all $x\in \Atom$.
\end{theorem}

\section{Relative truth and layered states}\label{sec:leyred}

Answers to a question can be either true or false. In the perspective of learning we think of truth values with respect to the actual knowledge that a learning agent can have at some stage of the process, so that we relativize the valuation of the answers to the knowledge states.
Furthermore the example of learning the solution to problem  (\ref{eq:non-monotonic}) in \S \ref{sec:learning} shows that there can be dependencies among answers in a state of knowledge. We formalize this by means of  the stratification into levels of the set of answers.
In the example of \S 1 we only need levels 0 and 1. In the definition, however, we allow any number of levels, even transfinite.

Denote with $\Ord$ the class of denumerable ordinals. Let us assume the existence of a map $\Level:\Atom \arrow \Ord$, associating to each answer  $x$ the
ordinal $\Level(x)$, and such that any two answers to the same question are of the same level. 
Thus, implicitly, the level is assigned to the questions.
If $X\in\InfState$ and $\alpha\in\Ord$ we write $X\restr \alpha = \Set{x\in X\mid \Level(x)<\alpha}$. We can now make precise the notions of level and of truth of an answer w.r.t. a knowledge state. Let us denote with $\Bool = \Set{\trueVal,\falseVal}$ the set of truth values.

\begin{definition}[Layered Knowledge Structures]\label{def:relTruth}
Let $(\Atom, \compat)$ be a knowledge structure. Then
a {\em layered knowledge structure} is a tuple $(\Atom, \compat, \Level, \Truth)$ where  $\Level:\Atom \arrow \Ord$ and $\Truth:\Atom \times \InfState \arrow \Bool$ are two maps such that:
\begin{enumerate}
\item {\em Two answers to the same question have the same level}:
\[ \forall x, y \in \Atom.
(x \compat y \Then \Level(x) = \Level(y))\]
\item
{\em $\Truth$ is continuous}, by taking $\Atom$ and $\Bool$ with the discrete topology, $\InfState$ with the state topology $\StateTop$, and $\Atom \times \InfState$ with the product topology;

\item  {\em $\Truth$ is layered:}
$\forall x \in \Atom, X \in \InfState. \;
\Truth(x,X) = \Truth(x,X\restr\Level(x))$

\end{enumerate}
\end{definition}
The ordinal $\Level(x)$ represents the reliability of the answer $x$, which is maximal when $\Level(x)=0$; by this reason we ask that all the answers to the same question have the same level. We prefer to assign levels to answers instead of questions because the formers are the only concrete objects of a knowledge structure.

The mapping $\Truth$ of a layered knowledge structure is a boolean valuation of atoms that continuously depends on states of knowledge; we call it a
{\em layered valuation} because the value of an atom w.r.t. a state (to which it does not necessarily belong) only depends on  the atoms of lower level belonging to that state.
Set $\True_x = \Truth(x)^{-1}(\Set{\trueVal}) = \Set{X \in \InfState \mid \Truth(x,X) = \trueVal}$, and similarly
$\False_x = \Truth(x)^{-1}(\Set{\falseVal})$.
When $\Truth$ is a layered valuation, if $X \in \True_x$ or $X \in \False_x$, we say that $x$ is true or false w.r.t. $X$ respectively. By definition, the truth of $x$ w.r.t. $X$ depends only on the \atoms of lower level than $\Level(x)$; it follows that
\[\Level(x) = 0 \Then \Truth(x,X) = \Truth(x,\emptyset)\]
that is, the truth value of \atoms of level $0$ is absolute, and it depends just on the choice of $\Truth$.  Level $0$ answers play the role of ``facts''. On the contrary answers of level greater than $0$ are better seen as empirical hypotheses, that are considered true as far as they are not falsified by answers of lower level.

\iniese{Continuing example \ref{ex:knowledge-structure}, let us set $\Level(i,n,m) = i$. Then we define
the meaning of the answers in $\Atom_2$ via the mapping $\Truth$ by putting: $\Truth((0,n,m), X) = \trueVal$ if and only if
$f_1(n) > f_1(m)$, and: $\Truth((1,n,m), X) = \trueVal$ if and only if
$n,m \in (\Nat_{f_1})^X$  and $f_2(n)>f_2(m)$. Recall that $(\Nat_{f_1})^X$ is the set of $n$ such that $(0,n,p) \not \in X$ for all $p$, and that if $(i,n,m) \in \Atom_2 = \Atom_0\uplus\Atom_1$ then $n < m$.

Clearly $\Truth$ is layered since $\Truth((i,n,m), X)$ does not depend on $X$ when $i=0$, while when $i = 1$ it depends on
$X\restr 1 = \Set{(i,n,m) \in X \mid i=0}$, which is morally $\Atom_0\cap X$.

To see that $\Truth$ is continuous let us observe that $\Truth((0,n,m), X)$ is the constant function w.r.t. $X$, and that
$\Truth((1,n,m), X) = \trueVal$ if and only if $f_2(n)>f_2(m)$ and
$n,m \in (\Nat_{f_1})^X$, and the set of $X$ for which this is true is a clopen as a consequence of Lemma
\ref{lem:complBxIsOpen}.
}\label{ex:layered-valuation}

\medskip
From now on, we assume that some layered knowledge structure $(\Atom, \compat, \Level, \Truth)$ has been fixed, with some level map $\Level$ and some continuous layered  valuation $\Truth$. We now introduce the set $\Sound$ of {\em sound} knowledge states (those from which nothing should be removed), the set $\Complete$ of {\em complete} knowledge states (those to which nothing should be added), the set $\Model$ of {\em model} states (the ``perfect'' states, those from which nothing should be removed and to which nothing should be added). The present notion of ``model'' is different and much weaker than the Tarskian's one; if atoms are to be logical statements, a ``model'' is a sound and complete theory; in general it defines the ideal subjective condition of having a full knowledge of the ``world'', namely of a model in the familiar sense of logic.

\begin{definition}[Sound and Complete States]\label{def:sounComplete}
Let $X\in \InfState$, $x \in \Atom$. Then:
\begin{enumerate}
\item $X$ is {\em sound} if $\forall x\in \Atom. \ x\in X \Then \Truth(x,X) = \trueVal$;
\item $X$ is {\em complete} if $\forall x\in \Atom. \ X\cap[x] = \emptyset \Then \Truth(x,X) = \falseVal$;
\item $X$ is a {\em model} if it is sound and complete.
\end{enumerate}
We call $\Sound$, $\Complete$ and $\Model$ the sets of sound, complete and model states respectively.
\end{definition}

In words a state of knowledge $X$ is sound if all the answers it contains are true w.r.t. $X$ itself; $X$ is complete if no answer which is true w.r.t. $X$ and compatible with the answers in $X$ can be consistently added to $X$; hence $X$ is a model if it is made of answers true w.r.t. $X$ and it is maximal. We think of a model $X$ as a perfect representation of the world.
For instance with respect to the examples in \S \ref{sec:learning} and \S \ref{sec:state_topology}, if $X$ is a model then the sets $(\Nat_{f_1})^X$ and $(\Nat_{f_1})_{f_2}^X$ are equal to $(\Nat_{f_1})$ and $(\Nat_{f_1})_{f_2}$ respectively, that is the beliefs of the agent perfectly agree with absolute truth.

In spite of this interpretation, models are far from being unique even w.r.t. a fixed map $\Truth$. Two models can include two different answers to the same question, because a question can have many true answers, while w.r.t. any state of knowledge each question is associated to a memory cell having room for a single answer.

Let us define $\Sound_x = \Set{X \in \InfState \mid x \in X \Then \Truth(x,X) = \trueVal}$, or equivalently
$\Sound_x = (\InfState \setminus A_x) \cup \True_x$; $\Complete_x = \Set{X \in \InfState \mid X\cap[x] = \emptyset \Then \Truth(x,X) = \falseVal}$, that is $\Complete_x = (\InfState \setminus B_x) \cup \False_x$, and $\Model_x = \Sound_x \cap \Complete_x$. Clearly we have $\Sound = \bigcap_{x \in \Atom} \Sound_x$, $\Complete = \bigcap_{x \in \Atom} \Complete_x$ and $\Model = \bigcap_{x \in \Atom} \Model_x$.

From a topological viewpoint, it is interesting to observe that all the above subsets of $\InfState$  are closed in $\StateTop$, while some of them are clopen.

\begin{lemma}
\label{lem:clopen}
For all $x \in \Atom$, $\True_x, \False_x, \Sound_x, \Complete_x, \Model_x$ are clopen in $\StateTop$.
$\Sound, \Complete, \Model$ are closed in $\StateTop$.
\end{lemma}

\proof For all $x\in\Atom$, the function $\Truth(x):\InfState \rightarrow \Bool$, where $\Truth(x)(X) = \Truth(x,X)$, is continuous. Then
by \ref{lem:complBxIsOpen}.\ref{lem:complBxIsOpen-iii}, and the fact that $\Bool$ is discrete, we deduce that $\True_x, \False_x$ are clopen. By \ref{lem:complBxIsOpen}.\ref{lem:complBxIsOpen-i} and \ref{lem:complBxIsOpen}.\ref{lem:complBxIsOpen-ii} the sets $B_x, A_x$ are clopen, therefore $\Sound_x, \Complete_x, \Model_x$ are clopen as well, being obtained by complement, finite unions and intersections of clopen sets. We conclude that $\Sound, \Complete, \Model$ are closed because they are intersections of closed sets.
\QED

\medskip
It is immediate that sound sets exist, as well as complete ones: trivial examples are $\emptyset$ which is vacuously sound, and
any set  $X$ including one answer $x$ for each equivalence class $[x]\in \Question$, which is vacuously complete, but not necessarily sound. Here is a non-trivial though simple example of these concepts.

\iniese{Suppose that $f_1(0) = 2 = f_1(n)$ for all $n>2$, and that $f_1(1) = 1$ and $f_1(2) = 0$. If we consider the states over
$\Atom_0$ only, and the restriction to $\Atom_0$ of the mapping $\Truth$ in Example \ref{ex:layered-valuation}, we have the models $\Set{(0,0,1),(0,1,2)}$ and $\Set{(0,0,2),(0,1,2)}$. Any subset of these sets is sound, while
$\Set{(0,n,n+1) \mid n \in\Nat}$ is complete but not sound.}\label{ex:soundCompleteModel}

\medskip
It is not obvious, however, that models exist in general. The proof that models exist is somehow reminiscent of G\"{o}del's proof that all first order consistent theories have a model.

\begin{theorem}[Existence of Models]\label{thr:models}
For every layered knowledge structure $(\Atom, \compat, \Level, \Truth)$
whose space of knowledge is $\InfState$, there exists a model $X\in \InfState$.
\end{theorem}

\proof Fix an arbitrary indexing $x_0, x_1, \ldots$ of the countable set $\Atom$.
For each $x\in \Atom$ and $Y\in \InfState$ set:
\[g(x,Y) = \left\{
		\begin{array}{cl}
		\Set{x_i} & \mbox{if $i$ is the minimum index $j$ s.t.} \\
						  & \mbox{$x_j \in [x] \wedge \Truth(x_j,Y) = \trueVal$, if it exists}
						  \vspace{2mm}\\
		\emptyset & \mbox{otherwise}
		\end{array}\right.\]
Now define inductively for each $\gamma\in \Ord$:
\[X_\gamma = \bigcup \Set{g(x,X_{<\gamma}) \mid \Level(x) = \gamma} ~~~~\mbox{where}~~~~
X_{<\gamma} = \bigcup_{\delta<\gamma} X_\delta \]
In words, $X_\gamma$ is obtained by choosing an answer
$x'$, if any, for each equivalence class $[x]$ with $\Level(x) = \gamma$, such that
$x'$ is true w.r.t. all the choices made at previous stages $\delta < \gamma$.
Since $x_i \in [x]$ implies that $\Level(x_i) = \Level(x)$, $X_\gamma$ is made of \atoms of level $\gamma$.

Then we prove that $X = \bigcup_{\gamma\in \SubOrd}X_\gamma$ is a model. First by construction
$X_\gamma$ is consistent for all $\gamma$, because it contains at most one answer for each equivalence class;
this implies that $X$ is consistent, since two \atoms in the same equivalence class are in the same $X_{\alpha}$.
Second, if $x\in X \restr \gamma$ then $x\in X_{<\gamma}$, so that:
\[\Truth(x,X) = \Truth(x,X\restr \Level(x)) = \Truth(x,X_{<\Level(x)}) = \trueVal\]
and $X$ is sound. Finally, for all $x \in \Atom$, if $X \cap [x] = \emptyset$ then
\[\begin{array}{llll}
X \cap [x] = \emptyset & \Then & X_{\Level(x)} \cap [x] = \emptyset \\
& \Then & \forall x' \in [x]. ~ \Truth(x', X_{<\Level(x)}) = \falseVal \\
& \Then & \Truth(x, X_{<\Level(x)}) =
\falseVal \\ & \Then & \Truth(x, X) = \falseVal
\end{array}\]
by $\Truth(x, X) =  \Truth(x, X_{<\Level(x)})$. Therefore $X$ is complete and hence a model.
\QED

The construction of Theorem \ref{thr:models} is not effective, even when the layered knowledge structure is recursive. Assume that $\gamma \in \Ord$ is the number of levels of the knowledge structure. If we look closely to the proof, we see that we defined a model by some  $\Delta^0_{1+\gamma}$-predicate. In particular, if there are infinitely many levels, then the definition is not an arithmetical predicate. We claim that the recursive complexity in our result is optimal: for any $\gamma$ there is some recursive layered knowledge structure with $\gamma$ levels, whose models are all  (the extensions of) $\Delta^0_{1+\gamma}$-complete predicates, and therefore are never $\Delta^0_{1+\delta}$-predicates, for any $\delta < \gamma$. In general models are not recursive sets, and {\em a fortiori} not finite.

\section{Interactive Realizability}\label{sec:realizer}

Given a layered knowledge structure $(\Atom, \compat, \Level, \Truth)$, the goal of a learning process is to reach some sound $X\in \InfState$ which is sufficiently large to compute a solution to the problem at hand, e.g. a $k$-tuple $n_1,\ldots,n_k$ of natural numbers witnessing the truth of (\ref{eq:monotonic}) or of (\ref{eq:non-monotonic}) in \S \ref{sec:learning}. To make this precise, we formally define what it means that a {\em problem} $P\subseteq C$,
(where $C$ is an arbitrary denumerable set, the carrier) has a solution $\alpha$ relative to a state $X$.
Informally, we require that $\alpha(X)$ is an element of $C$ continuously depending on a knowledge state $X$, which belongs to $P$
(or it satisfies the predicate $P$) whenever $X$ is a model. In the terminology of \cite{Berardi-deLiguoro:TOCL-11} $\alpha$ is an ``individual''.

We give definitions and results in full generality; however we are mainly interested in recursive $\alpha$ and $P$ and in effective ways of finding a state $X$ such that $\alpha(X)\in P$.

\begin{definition}[Solution of a Problem w.r.t. a Knowledge Structure]\label{def:validity}
Let $(\Atom, \compat, \Level, \Truth)$ be a layered knowledge structure and $\InfState$ its space of states of knowledge.
Given a continuous $\alpha:\InfState\arrow C$ where $C$ is taken with the discrete topology a predicate $P\subseteq C$ (a {\em problem}), and $X\in\InfState$, we define:
\begin{enumerate}
\item $X \models_{\Atom} \alpha:P \Iff \alpha(X)\in P$,
\item $\models_{\Atom} \alpha:P \Iff \forall X \in \InfState.\ X \ \mbox{is a model} \ \Then X \models_{\Atom} \alpha:P$.
\end{enumerate}
When  $\models_{\Atom} \alpha:P$ we say that $\alpha$ is a {\em solution} of $P$ w.r.t. $(\Atom,\compat)$.
\end{definition}
When $P$ is a predicate defined by a formula, the set $C$ is the carrier of a model in the usual sense;
therefore $\alpha$ is a multivalued individual akin to
individuals in a product model.
We shall omit the subscript $\Atom$ in $\models_{\Atom}$ when $\Atom$ is understood; in the examples we take $C = \Nat$.

\iniese{Let $(\Atom_2,\compat_2)$ be the knowledge structure defined in example \ref{ex:knowledge-structure} in \S \ref{sec:state_topology}, and $\InfState_2$ its knowledge space. Fix $k\in\Nat$; writing $\langle n_1,\ldots,n_k\rangle$ for the code number of the $k$-tuple $n_1,\ldots,n_k$ we define the ``problem'' $P_2$:
\[P_2 = \Set{\langle n_1,\ldots,n_k\rangle \mid  \bigwedge_{i < k} (n_i < n_{i+1} \ \wedge \ f_1(n_i) \leq f_1(n_{i+1}) \ \wedge
\   f_2(n_i) \leq f_2(n_{i+1})) },\]
$P_2$ is the set of all (coding of) $k$-tuple witnessing that $(\ref{eq:monotonic})$ and $(\ref{eq:non-monotonic})$ in \S \ref{sec:learning} are true.
Now for any $X\in\InfState$ define:
\[\alpha_2(X) = \min\Set{\langle n_1,\ldots,n_k\rangle \mid n_1 < \cdots < n_k \  \wedge \ n_1,\ldots, n_k \in (\Nat_{f_1})_{f_2}^X}\]
where $\min$ is understood as the lexicographic ordering of the $k$-tuples.
By definition the mapping $\alpha_2$ picks the first $k$ elements in the set $(\Nat_{f_1})_{f_2}^X$ in increasing order. Note that $\alpha_2$ is not a dummy search procedure, rather it is a primitive that reads of the guess for the solution to the problem $P_2$ given the knowledge $X$. $\alpha_2$ is always defined because $\Nat_{f_1}^X$ and $(\Nat_{f_1})_{f_2}^X$ are infinite for every $X\in\InfState_2$. Indeed this can be proved by a relativization to $X$ of the argument of Lemma \ref{lem:f-min}: if $n\not\in \Nat_{f_1}^X$ then there exists $m\in\Nat$ s.t. $n < m$ but $f_1(n) > f_1(m)$ in the knowledge state $X$, namely we have $(0,n,m) \in X$. Were $\Nat_{f_1}^X$ finite, we would be able to find infinitely many such $m$ forming an infinite increasing chain, and so an infinite descending chain via $f_1$. Similarly one proves that  $(\Nat_{f_1})_{f_2}^X$ is infinite (quantifying over $\Nat_{f_1}^X$ in place of $\Nat$ and coding the counterexamples known at $X$ by $(1,n,m) \in X$).

We show that $\alpha_2$ is continuous. Let $\alpha_2(X) = \langle n_1,\ldots,n_k\rangle$: then $n_i\in (\Nat_{f_1})_{f_2}^X$ for all $i\leq k$, and and for all $m< n_k$ with $m\neq n_1,\ldots,n_k$,  we have $m \not \in (\Nat_{f_1})_{f_2}^X$. Conversely one can check that for all $Y\in \InfState$, if $\bigwedge_{i\leq k} n_i\in (\Nat_{f_1})_{f_2}^Y$ and $\bigwedge_{m < n_k, m\neq n_1,\ldots,n_k} m \not \in (\Nat_{f_1})_{f_2}^Y$ then $\alpha_2(Y) = \langle n_1,\ldots,n_k\rangle$.
But since we know that the predicate $n \in  (\Nat_{f_1})_{f_2}^Y$ is continuous in $Y$ (see the remark after Lemma \ref{lem:complBxIsOpen}), the last condition defines a finite intersection of clopens, which is clopen.

We show that $\alpha_2$ is a solution of $P_2$, that is, that $\models \alpha_2:P_2$. Let $X$ be a model: then we have $(\Nat_{f_1})_{f_2}^X = (\Nat_{f_1})_{f_2}$. Since
$\alpha(X) = \langle n_1,\ldots,n_k\rangle \in (\Nat_{f_1})_{f_2}^X$, we deduce $\alpha_2(X)\in (\Nat_{f_1})_{f_2}$, that is, that
$f_1$ and $f_2$ are weakly increasing w.r.t. $n_1,\ldots,n_k$. Thus, $\alpha_2(X)\in P_2$.
}\label{ex:solution}

\medskip
A solution $\alpha$ is some way to produce an inhabitant $\alpha(X) \in P$ out of any model $X$.
A learning strategy for a problem $P$ admitting a solution $\alpha$ w.r.t. $(\Atom,\compat,  \Level, \Truth)$ is ideally a search procedure of some model $X\in\InfState$. But models are in general infinite and non-recursive states of knowledge: to make learning effective we rely on the continuity of $\alpha$ which implies that if $\alpha(X) = n \in P$ for some model $X$ there exists a finite $s\subseteq X$ such that
$\alpha(s) = n$.

We describe the search of such finite sound approximations of a model $X$ via certain continuous functions $r$ over $\InfState$. The function $r$ is such that for any sound $X$ (not necessarily a model) the set $r(X)$ is a finite set of answers that are not in $X$ but are compatible with the answers in $X$ and true w.r.t. $X$. So $r$ is a mean to extend a state of knowledge with new answers; when $r(X) = \emptyset$ there is nothing else to know and we expect that $\alpha(X) \in P$. When such a function exists for given $P$ and $\alpha$ we say that it is a {\em realizer} of the statement that $\alpha$ is a solution of $P$. The name ``realizer'' comes from \cite{Berdel08,AschBerLMCS2010,Berardi-deLiguoro:TOCL-11}; in particular in \cite{AschBerLMCS2010} it is shown how Kleene's
realizability can be extended to HA plus a restricted form of excluded middle by considering
as realizers functionals depending on  ``states'', that are states of knowledge with just one level. Indeed
the present concept subsumes that of realizers considered in the previous works.
\newpage

\begin{definition}[Realizers and Zeros]\label{def:realizer}
A {\em realizer} is a continuous map $r:\InfState\arrow\FinPow(\Atom)$ such that for all $X\in\InfState$ and all $x\in r(X)$:
\begin{enumerate}
\item \label{def:realizer-i}  $X \cap [x] = \emptyset$,
\item \label{def:realizer-ii} $\Truth(x,X) = \trueVal$.
\end{enumerate}
We denote by $\Realizer$ the set of realizers. Finally we say that
$X \in \InfState$ is a {\em zero} of $r\in\Realizer$ if $r(X) = \emptyset$.
\end{definition}
For the sake of generality we do not assume any condition about the $X$ above: in particular it is not necessarily sound.

We can see a realizer $r$ as the essential part of a learning strategy, which tries to update the current state of knowledge. This is obtained by evaluating $r(X)$ to get a finite set of new answers by which $X$ could be soundly extended.
To see this let $\New:\InfState \times \FinPow(\Atom) \arrow \FinPow(\Atom)$ be defined  by:
\[\New(X,U) = \Set{x\in U \mid X \cap [x] = \emptyset \wedge \Truth(x,X) = \trueVal}.\]
In words $\New(X,U)$ selects form $U$ those atoms that can be consistently added to the state of knowledge $X$.

\begin{lemma}\label{lem:newContinuous}
The function $\New$ is continuous, where $\InfState$, $\FinPow(\Atom)$ and $\InfState \times \FinPow(\Atom)$ are taken with $\StateTop$,
the discrete and the product topology respectively.
\end{lemma}

\hide{
\proof
Since $\FinPow(\Atom)$ has the discrete topology, it suffices to show that the set $\Set{X \in \InfState \mid \New(X,U) = V}$ is open in $\StateTop$ for all $U,V \in \FinPow(\Atom)$. By definition of $\New$ for all $x \in U$, all $Y \in \InfState$ we have $\New(Y,\Set{x}) = \Set{x}$ if and only if $Y \in \InfState \setminus \Complete_x$, and $\New(Y,\Set{x}) = \emptyset$ if and only if $X \in \Complete_x$. Let $\Open = (\bigcap_{x \in V} (\InfState \setminus \Complete_x)) \cap (\bigcap_{x \in (U \setminus V)} \Complete_x) $. Then $\Open$ is a clopen, because it is obtained by complement and finite intersections of the sets $\Complete_x$, which are clopen by Lemma \ref{lem:clopen}. By construction, $Y \in  \Open$  if and only if $\New(Y, U) = V$.
\QED
}

Say that an {\em operator} over  $\InfState$ is any  continuous map $r:\InfState\arrow\FinPow(\Atom)$, and call $\Operator_{\InfState}$, or simply $\Operator$, the set of operators over $\InfState$. For $r\in\Operator$ we set:
$\realz{r}(X) = \New(X,r(X))$.
%Note that the set of realizers is the subset of operators such that
%$r = \realz{r}$.

\hide{
\begin{lemma}\label{prop:overRealiz}
The set of realizers is the subset of operators such that
$r = \realz{r}$.
\end{lemma}

\Proof By Lemma \ref{lem:newContinuous} and the fact that $\realz{r} = \New \comp \Pair{\Id_{\InfState}}{r}$, $\realz{r}$ is continuous as soon as $r$ is such. On the other hand by the very definition:
\[\realz{r}(X) = \New(X,r(X)) = \Set{x\in r(X) \mid X\cap [x] = \emptyset \wedge \Truth(x,X)}\]
Therefore $\realz{r} \in \Realizer$. Beside if $r\in\Realizer$ then condition $X\cap [x] = \emptyset \wedge \Truth(x,X)$ holds for all $x\in r(X)$, hence $\realz{r} = r$. From this and the fact that $\realz{r}\in \Realizer$ for all $r\in\Operator$ it follows that $\realz{\realz{r}} = \realz{r}$ for all operator $r$, and we are done.
\QED
}

\iniese{We propose a realizer solving the problem (\ref{eq:monotonic}), expressed by the predicate:
\[P_1 = \Set{\langle n_1,\ldots,n_k\rangle \mid  \bigwedge_{i < k} (n_i < n_{i+1} \ \wedge \ f_1(n_i) \leq f_1(n_{i+1})
) }.\]
We first define a function $\beta_1(X, \langle n_1,\ldots,n_h \rangle)$ extending any list $\langle n_1,\ldots,n_h\rangle$ with $h \le k$ to a solution of (\ref{eq:monotonic}):

\[\beta_1(X, \langle n_1,\ldots,n_h\rangle) =
\left\{ \begin{array}{ll}
	\langle n_1,\ldots,n_h \rangle &  \mbox{if $h= k$} \\ [1mm]
	\beta_1(X, \langle n_1,\ldots,n_h, m \rangle) & \mbox{where $m$ is the minimum s.t.} \\
	& \mbox{$m\in \Nat_{f_1}^X$ and} \\
	& \mbox{$m>n_h$ if $h>0$.}
	\end{array}
\right.\]
A function solving (\ref{eq:monotonic}) may then be defined by $\alpha_1(X) = \beta_1(X, \langle \, \rangle)$. We claim that $\models\alpha_1:P_1$.
Indeed $\beta_1$ is continuous w.r.t. $X$ since $m \in \Nat_{f_1}^X$ is equivalent
to $X\not\in B_{(0,m,m+1)}$ which is a clopen by \ref{lem:complBxIsOpen}.\ref{lem:complBxIsOpen-i}, so that $\alpha_1$ is continuous. Further, if $X$ is a model, then $\alpha_1(X)\in P_1$.
We define a realizer $r_1$ looking for some $X$ such that $\alpha_1(X)$ solves $P_1$. $r_1(X)$ takes any knowledge state $X$, and adds to it the first counterexample to (\ref{eq:monotonic}) we may find in the list $\langle n_1, \ldots, n_k \rangle$ generated by $\alpha_1$, unless $\alpha_1(X)$ solves $P_1$. We define $r_1$ in two steps: first, we define a map $g_1$ finding the first counterexample to (\ref{eq:monotonic}) in a given list:
\[g_1(\langle n_1,\ldots,n_k \rangle) = \left\{\begin{array}{ll}
\{(0,n_i, n_{i+1}) \mid \mbox{$1\leq i< k$ min. s.t. } \\
\hspace{2.5cm} \mbox{$f_1(n_i)>f_1(n_{i+1})$}\} & \mbox{if $i$ exists} \\[1mm]
\emptyset & \mbox{otherwise.}
\end{array}
\right.\]
Then we define the realizer by composing $g_1$ with the output of $\alpha_1$: $r_1(X) = g_1(\alpha_1(X))$. $r_1$ is a realizer, because $r_1(X)$ always outputs atoms not in $X$. Indeed, if $\alpha_1(X) = \langle n_1,\ldots,n_k \rangle \in \Nat_{f_1}^X$, and if $r_1(X)$ output the atom $f_1(n_i)>f_1(n_{i+1})$, then $(f_1(n_i)>f_1(n_{i+1})) \not \in X$ by definition of $\langle n_1,\ldots,n_k \rangle \in \Nat_{f_1}^X$. We may use $r_1(X)$ to extend $X$ until we find some $X$ such that $r_1(X) = \emptyset$. Whenever $r_1(X) = \emptyset$ we have $g_1(\alpha_1(X)) = \emptyset$, hence  $\alpha_1(X)$ solves (\ref{eq:monotonic}) by definition of $g_1$.

To step from $P_1$ to $P_2$, namely to problem (\ref{eq:non-monotonic}), we just replace $\Nat_{f_1}^X$ by $\Nat_{f_1,f_2}^X$, namely:
\[\beta_2(X, \langle n_1,\ldots,n_h\rangle, k) =
\left\{ \begin{array}{ll}
	\langle n_1,\ldots,n_h \rangle &  \mbox{if $h= k$} \\ [1mm]
	\beta_2(X, \langle n_1,\ldots,n_h, m \rangle, k) & \mbox{where $m$ is the minimum s.t.} \\
	& \mbox{$m\not\in (\Nat_{f_1})_{f_2}^X$} \\
	& \mbox{and $m>n_h$ if $h>0$.}
	\end{array}
\right.\]
and
\[g_2(\langle n_1,\ldots,n_k \rangle) = \left\{\begin{array}{ll}
\{(0,n_i, n_{i+1}) \mid \mbox{$1\leq i< k$ min. s.t. } \\
\hspace{0.5cm} \mbox{$f_1(n_i)>f_1(n_{i+1})\vee f_2(n_i)>f_2(n_{i+1})$}\} & \mbox{if $i$ exists} \\[1mm]
\emptyset & \mbox{otherwise.}
\end{array}
\right.\]
We have that $\alpha_2(X) = \beta_2(X, \langle \, \rangle, k)$, where $\alpha_2$ is from example \ref{ex:solution}. Now let us
define $r_2(X) = g_2(\beta_2(X, \langle \, \rangle, k)) = g_2(\alpha_2(X))$. Then
we can show that $r_2$ is a realizer looking for some $X$ such that $\alpha_2(X) \in P_2$ just as in the case of $r_1$ above.
}\label{ex:realizer}

\bigskip
If $X$ is a model and $r\in\Realizer$ then $r(X) = \emptyset$ by definition;
on the other hand if  $\models_{\Atom} \alpha:P$ then $\alpha(s) = \alpha(X) = n \in P$ for some finite $s\subseteq X$. Since $r$ is continuous, the condition that reveals that the approximation $s$ of $X$ is good enough to compute an $n \in P$ is that
$r(s) = r(X) = \emptyset$. This suggests that a constructive way to meet the requirement about models in the definition of
$\models_{\Atom} \alpha:P$ is to ask that sound zeros of a realizer $r$ are enough to find inhabitants of $P$ via $\alpha$, and then look for finite sound zeros of $r$. We turn this into the following definition.
\newpage

\begin{definition}[Interactive Realizability]\label{def:realPred}
Let $\alpha:\InfState\arrow C$ be continuous (w.r.t. the discrete topology over $\Nat$), and  $P\subseteq C$ a predicate over the carrier $C$:
\begin{enumerate}
\item $r\in\Realizer$ {\em interactively realizes} $P$ w.r.t. $\alpha$, written $r \real \alpha:P$, if and only if:
	\[\forall X \in \InfState. \; \mbox{$X$ sound zero of $r$} \Then \alpha(X) \in P\]
\item $P$ is {\em interactively realizable} w.r.t. $\alpha$, written $\real \alpha:P$, if and only if:
	\[\exists r\in\Realizer. \; r \real \alpha:P.\]
\end{enumerate}
\end{definition}
If $P_1$ and $P_2$ are the predicates with carrier $\Nat$ defined in examples \ref{ex:solution} and \ref{ex:realizer}, $\alpha_1,\alpha_2$ their respective solutions and $r_1,r_2$ the realizers from example \ref{ex:realizer}; then it can be proved that
$r_i \real \alpha_i:P_i$ for both $i=1,2$.

The main result of the paper is that the apparently stronger $r \real \alpha:P$ for some $r\in\Realizer$ is equivalent to
$\models\alpha:P$. That is, whenever $\alpha:P$ is valid then it is interactively learnable, and we have some strategy to find some finite $X$ such that $\alpha(X) \in P$.

Before we establish the existence of sound finite zeros of any $r\in\Realizer$. This is a non trivial fact because, whenever we add to some state $Y$ (no matter whether finite or not) a $y\in r(Y)$, we know that $\Truth(z,Y)=\trueVal$ for all $z\in Y$,
but we do not know about the value of $\Truth(z,Y\cup \Set{y})$, so that $Y\cup \Set{y}$ is not necessarily sound.
Moreover it is not true that if $s\subseteq X$ and $X$ is sound then $s$ is sound.

\iniese{Let us redefine $f_1$ by $f_1(0) = 10, f_1(1) = 30, f_1(2) = 20$ and define $f_2(0) = 20, f_2(1)=10, f_2(2)=20$. Also we let
$x = (1,0,1)$ meaning that $0\not\in (\Nat_{f_1})_{f_2}$, and $y = (0,1,2)$ meaning that $1\not\in \Nat_{f_1}$. Then $\Truth(x,\Set{x})=\trueVal$ since at $\Set{x}$ it is likely that $0\not\in (\Nat_{f_1})_{f_2}$ because of the counterexample in the point $1$; but
$\Truth(x,\Set{x,y})=\falseVal$ because the discovery that $1\in\Nat_{f_1}$ contradicts counterexample on point $1$.
}

\medskip
We prove below that finite sound zeros exist for all $r\in\Realizer$ and that these are finite approximations of sound states of knowledge which are themselves zeros of $r$, hence in particular of models.

\begin{lemma}\label{lem:soundExtension}
If $X\in \InfState$ is sound, $s\in \State$ is a finite state such that $s\subseteq X$, then there exists a finite sound $t \in \State$ such that $s \subseteq t \subseteq X$.
\end{lemma}

\hide{
\proof First let us extend the definition of $\Level$ to (finite) states: $\Level(s) = \max\Set{\Level(x) \mid x \in s}$ if $s\neq \emptyset$, $\Level(\emptyset) = 0$. Now by induction over $\Level(s)$, we prove that if $s \subseteq X$,
then there exists $t \subseteq X$ which is finite, sound and such that $s \subseteq t$; moreover $t$ can be chosen such that $\Level(t) = \Level(s)$.

If $\Level(s) = 0$ then for all $x \in s$ it is the case that
\[\Truth(x, s) = \Truth(x,\emptyset) = \Truth(x,X) = \trueVal\]
because of $\Level(x) = 0$. Hence in this case $t = s$.

Let $\Level(s) = \alpha > 0$: then let $s' = \Set{x \in s \mid \Level(x) = \alpha}$ and $s'' = s \setminus s'$.
If $s'' = \emptyset$ then $\Level(s'') = 0 < \alpha$ by definition; otherwise $\Level(s'') = \max\Set{\Level(y)\mid y\in s''} < \alpha$ by construction.

Suppose that $s' = \Set{x_1,\ldots,x_k}$: since $s' \subseteq s \subseteq X$ we know that
\[\Truth(x_i, X) = \Truth(x_i, X\restr \alpha) = \trueVal\] for all $1 \leq i \leq k$.
By continuity, for all $i$ there exists a basic open $\Open_{s_i,t_i}$ s.t. $X \restr \alpha \in \Open_{s_i,t_i}$
and $\Truth(x_i, \Open_{s_i,t_i}) = \trueVal$.
It follows that $\Open = \Open_{s_1,t_1} \cap \cdots \cap \Open_{s_k,t_k}$ is open, $X \restr \alpha \in \Open$ and
$\Truth(x_i,\Open) = \trueVal$ for all $i$. Let $U = s_1 \cup \cdots \cup s_k \subseteq X \restr \alpha$ and $V = t_1 \cup \cdots \cup t_k$; since $X\restr\alpha \in\Open$, we know that $( X\restr\alpha ) \cap V = \emptyset$. Thus, for any $t$ s.t. $U \subseteq t \subseteq X \restr\alpha$ it has to be the case that $t\in\Open$, and therefore
$\Truth(x_i,t) = \trueVal$ for all $i$.

Let $W = U \cup s'' \subseteq X \restr \alpha$, so that $\Level(W) = \beta < \alpha$, because $U\subseteq X\restr\alpha$.
By induction there exists a sound and finite $t'$
such that $t' \subseteq X$, $\Level(t') = \beta$ and $W\subseteq t' \subseteq X \restr \alpha$.

By construction $s'' \subseteq t'$, so that $s \subseteq t' \cup s' \subseteq X$ and $\Level(t' \cup s') = \alpha$. On the other hand $t' \cup s'$ is sound, since for all $y \in t' \cup s'$, either $y \in t'$, and then
\[\Truth(y, t' \cup s') = \Truth(y, t') = \trueVal\]
because $\Level(s') = \alpha > \Level(t') \geq \Level(y)$ and $t'$ is sound. Otherwise, if $y\in s'$ we have
that $\Level(y) = \alpha = \Level(s')$ so that
\[\Truth(y, t' \cup s') = \Truth(y, (t' \cup s')\restr\alpha) = \Truth(y, t')\]
But since $t'\in\Open$ by the above, we conclude that $\Truth(y, t') = \trueVal$. \QED
}

We are now in place to conclude the proof that every realizer has a finite sound zero.

\begin{theorem}[Existence of Sound and Finite Zeros of Realizers]\label{thr:soundFiniteZero}
If $r \in \Realizer$, then there exists a finite sound zero $t \in \State$ of $r$.
\end{theorem}

\proof Models exist by Theorem  \ref{thr:models}  and they are sound by definition, hence $r(X) = \emptyset$ for some sound $X \in \InfState$ since $r\in\Realizer$. By continuity there is a basic open $\Open_{s_0,t_0}$ such that $X \in \Open_{s_0,t_0}$ and $r(\Open_{s_0,t_0}) = \emptyset$. This implies that $s_0 \subseteq X$ and $X \cap t_0 = \emptyset$, so that a fortiori any finite $t \in \State$ such that $s_0 \subseteq t \subseteq X$ satisfies $t \cap t_0 = \emptyset$ and therefore $t \in \Open_{s_0,t_0}$, i.e. it is a zero of $r$. By Lemma \ref{lem:soundExtension} there exists a sound $t$ among them, which is the desired finite sound zero of $r$.\qed

We come now to the completeness theorem. Our thesis is that interactive realizability is complete in the sense that if $\alpha(X) \in P$ for all models $X$, then we may replace the model $X$ by the finite sound zeros of a suitable realizer $r \in \Realizer$. The proof of completeness is somehow reminiscent of G\"{o}del's proof that valid formulas are provable, with realizers in the place of classical proofs.

\begin{theorem} [Completeness of Realization]
\label{thr:sound-complete}
For any continuous $\alpha:\InfState\arrow C$ and predicate                                                                                                                                                                                                                                                                                                                  $P\subseteq C$ over the carrier $C$:
\begin{enumerate}
\item \label{thr:sound-complete-i} $\real \alpha:P$ if and only if $ \models \alpha:P.$
\item \label{thr:sound-complete-ii} If $ \models \alpha:P$ then the realizer
	$r$ such that  $r \real \alpha:P$  can be chosen recursive relatively to $\alpha$ and $P$.
\end{enumerate}                                                                                                                                                                                                                                                                                                                       \end{theorem}
                                                                                                                                                                                                                                                                                                                                                           \proof   We prove  (\ref{thr:sound-complete-i}) and  (\ref{thr:sound-complete-ii})    simultaneously.
$(\Then)$ If $X\in\InfState$ is a model then $X$ is a sound zero of any realizer by Definition \ref{def:realizer}; hence if $r \real \alpha:P$ for some $r\in\Realizer$ we immediately have $\alpha(X) \in P$, i.e. $X \models \alpha:P$ for arbitrary model $X$.

$(\Leftarrow)$
We have to show that, if $\;\models \alpha:P$, namely if $\alpha(X)\in P$ for $X \in \Model$, then there exists an $r\in\Realizer$
recursive in $\alpha$ and $P$,
such that $r \real
\alpha:P$. We establish the contrapositive:
	\[\alpha(X) \not\in P \Then \mbox{$X$ not sound} \ \vee
\ r(X) \neq \emptyset\]
	for some realizer $r$ and arbitrary $X\in\InfState$.

If $\alpha(X) \not\in P$ then, by the hypothesis, $X \not \in \Model$, hence $X \not \in \Sound$ or $X \not \in \Complete$. By definition of $\Sound$ and $\Complete$, this implies that $\exists x\in\Atom.\; X \not \in \Sound_x \vee X \not \in \Complete_x$. Fix an enumeration $x_0, x_1, \ldots$ of the countable set $\Atom$. Let us define $r:\InfState\arrow \FinPow(\Atom)$ recursive in $\alpha$ and $P$:
	\[r(X) = \left\{\begin{array}{ll}
				\emptyset & \mbox{if $\alpha(X)\in P$} \\
[1mm]
\Set{x_i} & \mbox{where $i = \min\Set{j \in\Nat \mid X\in \InfState \setminus \Model_{x_j}}$, else.}
				\end{array}\right.\]
Then $r$ is a total function since if $\alpha(X)\not\in P$ then $X \in \InfState - \Model$ so that $\Set{x_j \in\Atom \mid X \in \InfState \setminus \Model_{x_j}}\neq\emptyset$. If $r$ is continuous then $\realz{r}(X) = \New(X,r(X))$ because $X \in \InfState \setminus \Model_{x_j}$ implies $X \in \InfState \setminus \Complete_{x_j}$. Thus, $r$ is a realizer by Prop. \ref{prop:overRealiz}. We have $\realz{r} \real \alpha:P$. Indeed, assume for contradiction that $\alpha(X) \not \in P$, $X \in \Sound$ and $\realz{r}(X) = \emptyset$. Then $r(X) = \Set{x_i}$ and $X \in \InfState \setminus \Model_{x_i} = (\InfState \setminus \Sound_{x_i}) \cup (\InfState \setminus \Complete_{x_i})$. Since $X \in \Sound \subseteq \Sound_{x_i}$, then $X \in \InfState \setminus \Complete_{x_i}$. We conclude that $\realz{r}(X) = \Set{x_i }$, contradiction.

To see that $r$ is continuous it suffices to check that both $r^{-1}(\emptyset)$ and $r^{-1}(\Set{x})$ (for any $x\in\Atom$) are opens in $\StateTop$. Now  $r(X) = \emptyset$ if and only if $\alpha(X) \in P$, that is $X\in\alpha^{-1}(P)$ which is clopen by Lemma \ref{lem:clopen}.
On the other hand $X \in
r^{-1}(\Set{x})$ if and only if:
	 \[\exists i. \ x_i = x \ \wedge \  X \in (\InfState \setminus \Model_{x_i})\wedge \
\forall j < i.\, X \in \Model_{x_j}. \]
This is equivalent to $X \in \Model_{x_0} \cap \ldots \cap \Model_{x_{i-1}} \cap (\InfState \setminus \Model_{x_i}) $ which, by Lemma \ref{lem:clopen}, is a finite intersection of clopens, hence a clopen itself. \qed

\section{An algorithm to compute finite sound zeros of realizers}\label{sec:algorithm}

Theorem \ref{thr:soundFiniteZero} establishes that any realizer has a finite sound zero, which can be found by blind search. In this section we address the question of a realistic computation  of one zero given an effective realizer over an effective knowledge structure.

Let $(\Atom, \compat, \Level, \Truth)$ be a layered knowledge structure, and assume for simplicity that the image of $\Atom$ by
$\Level$ is just the ordinal $2$, namely that there are only two levels. Let $\InfState$ and $\State$ denote the space of knowledge states and of finite knowledge states over $\Atom$ respectively.
Recall that given an answer $x\in\Atom$ we say that it is a {\em fact} if $\Level(x) = 0$, and a  {\em hypothesis}
if $\Level(x) = 1$.

Suppose that $\Atom$ and $\compat$ are decidable, and $\Level$ and $\Truth$ are computable. Suppose also that $r \in \Realizer$ is a realizer for this structure, and that it is computable.
To compute a (finite) sound zero of $r$ a tentative algorithm, generalizing what we have seen in \S \ref{sec:learning}, is as follows:

\begin{tabbing}
$X := $ any finite sound state of knowledge, e.g. $\emptyset$ \\ [1mm]
while \= $r(X) \neq \emptyset$  let $x\in r(X)$  \\ [1mm]
          \> $Y := \Set{x\in X \mid \Truth(x,X\cup \Set{x}) = \trueVal}$ \\ [1mm]
          \> $X := Y \cup  \Set{x}$\\
return $X$
\end{tabbing}

\noindent The idea is to compute the new answers $r(X)$ that are compatible and true at $X$, and to choose non deterministically some $x \in r(X)$, whenever it exists, and to add $x$ to $X$
until we reach some $X'$ such that nothing can be added to $X'$ (that is such that $r(X') = \emptyset$). The intermediate step of computing $Y \subseteq X$ is needed
because if we add $x$ we might falsify some hypothesis $y \in X$ (i.e., some $y$ of level $1$),
and therefore we have to remove it.
This is why the subsequent values of $X$ do not form a chain w.r.t. subset inclusion in general, and we dub the whole process  ``non-monotonic learning''.

However the tentative algorithm does not always terminate: this is due to the fact that,
before being removed, $y$ could be used by $r$ to add some new hypothesis $y'$ to $X$,
then $y'$ is removed only after $r$ uses $y'$ to add some new hypothesis $y''$ to $X$, and so forth. The only way to prevent divergence is to remove $y$ together with all the hypothesis that have been added in any previous step {\em because of $y$}.
For the sake of simplicity we add an integer variable to keep track of ``time'', and remove from $X$ all the $z$ that have been added after $y$, which include all those logically depending on $y$, thought do not necessarily coincide with them. By using simple sets of pairs $(x, n)$ where $x\in\Atom$ and $n\in\Nat$ we record at what step an answer has been added to the current state of knowledge, and  we use this information while removing hypothesis from the state $X$. The algorithm is modified as follows:

\medskip
\noindent Algorithm {\bf Find Sound Zero}:
\begin{tabbing}
1. \= input any finite sound state of knowledge $X\in \State$ and computable $r\in\Realizer$ \\ [1mm]
2. \> $D := \Set{(x, 0) \mid x \in X}$  and $n := 0$ \\ [1mm]
3. \> while \= $r(X) \neq \emptyset$  let $x \in r(X)$ \\ [1mm]
4. \>           \> $E := \Set{(y, m) \in A \mid
          	\neg \exists\, (z,k) \in D. \; \Level(z) \leq \Level(y) \wedge k \leq m \wedge
				 \Truth(z,X\cup \Set{x}) = \falseVal  }$ \\ [1mm]
5. \>           \> $D := E \cup  \Set{(x, n)}$ and $n := n+1$\\ [1mm]
6. \>           \> $X := \Set{z \in \Atom \mid \exists \, i \in \Nat.\; (z,i) \in D}$ \\ [1mm]
7. \>  return $X$
\end{tabbing}
\bigskip

\noindent
Let $X_0, X_1, \ldots $ be the subsequent values of the program variable $X$; then we claim that the following is an invariant of the algorithm loop:

\begin{lemma}\label{lem:invariant}
 For all $i$,  $X_i \in \State$ and it is sound.
\end{lemma}

\proof
The thesis holds for $X_0$ by assumption. Assume that $X_i$ is a finite sound state of knowledge and consider $X_{i+1}$.
Because of the assignment in step $4.$ we know that $E \subseteq D$, so that by the assignments in steps $5.$ and $6.$,
$X_{i+1} \subseteq X_i \cup \Set{x}$.
By the fact that $r$ is a realizer we know that $X_i \cap [x] = \emptyset$, and therefore $X_i \cup \Set{x} \in \State$, so that
$X_{i+1} \in \State$.

To see that $X_{i+1}$ is sound, let us observe that at step $4.$ a pair $(z,k)$ cannot be removed from $D$ if $\Level(z) = 0$,
because $\Truth(z,X_i) = \trueVal$ by the soundness of $X_i$ and
$\Truth(z,X_i \cup \Set{x}) = \Truth(z,(X_i \cup \Set{x})\restr 0) = \Truth(z,\emptyset) = \Truth(z,X_i) = \trueVal$, because
in that case $z$ is a fact. Consider now the case of $z \in X_{i+1}$ with  $\Level(z) = 1$, namely when $z$ is a hypothesis.
Since we know that  $X_{i+1} \subseteq X_i \cup \Set{x}$, we have that $\Truth(z,X_i) = \trueVal$, either
because $z \in X_i$ which is sound, or because $z = x$ and $r$ is a realizer.
If $\Level(x) = 1$ then by the above remark that no atom of level $0$ can be removed from $X_i$, we have
that $X_{i+1}\restr 1 = X_i\restr 1$ which implies that $\Truth(z,X_{i+1}) = \Truth(z,X_{i+1}\restr 1) = \Truth(z,X_i\restr 1) =
\Truth(z,X_i) = \trueVal$. Therefore the only interesting case is when $\Level(z) = 1$ and $\Level(x) = 0$ where $x$ is the atom
newly chosen in $r(X_i)$, because this is right the case in which the valuation of a hypothesis might change. However we observe that in this case $X_{i+1}\restr 1 = X_i\restr 1 \cup \Set{x}$, and that if $z\in X_{i+1}$ then $(z,m) \in E$ for some $m$; it follows
that $\Truth(z, X_{i+1}) = \Truth(z, X_{i+1}\restr 1) = \Truth(z, X_i\restr 1 \cup \Set{x}) = \Truth(z,X_i \cup \Set{x}) \neq \falseVal$ by the condition in step $4.$ and hence $\Truth(z, X_{i+1}) = \trueVal$. \qed

\medskip
It remains to show that the above algorithm always terminates. Before embarking into the proof let us recall that, in general, the
sequence $X_0, X_1, \ldots $ is not increasing w.r.t. subset inclusion, because hypothesis appearing in the earlier stages can be
removed at step $4.$ from subsequent ones. Nonetheless we have that $X_0\restr 1 \subseteq X_1\restr 1 \subseteq \cdots $,
so that by the continuity of the realizer $r$ we can argue that the values of $r$ over that chain are eventually constant:

\begin{lemma}\label{lem:stab_mon}
If $Y_0 \subseteq Y_1\subseteq \cdots$ is a non decreasing chain of states in $\InfState$ and $r\in\Realizer$ then there is an $n_0$ such that for all $n \geq n_0$, $r(Y_n) = r(Y_{n_0})$.
\end{lemma}

\proof Let $Y = \bigcup_n Y_n$. Then $Y\in \InfState$ since, by the hypothesis that  $Y_0 \subseteq Y_1\subseteq \cdots$, any  $x,y \in Y$  belong to some $Y_n$ which implies that either $x = y$ or $x \not\compat y$ (that is $x \# y$). By continuity of $r$ there is a basic open $\Open_{U,V}  = \bigcap_{x\in U} A_x \cap\, \bigcap_{y\in V} B_y$, with finite $U,V$, such that $Y\in \Open_{U,V}$ and $r(\Open_{U,V}) = r(Y)$. The hypothesis that $\Set{Y_i}$ is a chain and that $U$ is finite
implies that there is $n_0$ such that for all $n\geq n_0$, $U\subseteq Y_n$; on the other hand, for all $z\in V$, it is the case that $Y \cap [z] = \emptyset$ so that
$Y_n \cap [z] = \emptyset$ since $Y_n \subseteq Y$. We conclude that
for all $n\geq n_0$ it is $Y_n\in \Open_{U,V}$ and therefore $r(Y_n) = r(Y) = r(Y_{n_0})$. \qed

As observed above, the $X_i$ do not form a chain w.r.t. subset inclusion; however we can extract a subsequence of $X_0,X_1,\ldots$ such that the
monotonicity condition does hold.
Let $D_i$ and $E_i$ be the values of $D$ and $E$ after $i$ steps of the algorithm respectively. In case $r(X_i) = \emptyset$ we set $X_j = X_i$, $D_j = D_i$ and $E_j = E_i$ for all $j > i$, so that the $X_i$, $D_i$ and $E_i$ are always infinite sequences of (finite) sets, and if $r(X_i) = \emptyset$ for some $i$ then
the sequences are definitely constant and the sets $r(X_j)$ are definitely empty.

Let $D_\infty = \Set{(x,k) \mid \forall i \geq k.\; (x,k) \in D_i}$, namely the set of pairs $(x,k)$ which are definitely elements of the $D_i$. Also consider the statements:
\[(\mbox{Eq}_i): \hspace{8mm} E_i = \Set{(x,k) \in D_\infty\mid k < i}.\]
Then we prove in two steps that the equations $(\mbox{Eq}_i)$ are true for infinitely many $i$.

\begin{lemma}\label{lem:Eqi-inclusione}
For all $i$ we have $\Set{(x,k) \in D_\infty\mid k < i} \subseteq E_i$.
\end{lemma}

\proof If $(x,k) \in D_\infty$ then $(x,k) \in D_k$; if $k < i$ but $(x,k) \not\in E_i$ then $(x,k) \not\in D_i$ since $D_i = E_i \cup \Set{(z,i)}$ by the assignment $5.$
and even if $z = x$ it will be $i \neq k$. By this we get the contradiction $(x,k) \not\in D_\infty$. \qed

\begin{lemma}\label{lem:inclusione_infOften}
For all $i$ there is $j \geq i$ such that $(\mbox{\rm Eq}_j)$ holds.
\end{lemma}

\proof %By Lemma \ref{lem:Eqi-inclusione} it suffices to prove that $E_i \subseteq \Set{(x,k) \in D_\infty\mid k < i}$ infinitely often.
Let $\Delta_i = E_i\setminus\Set{(x,k) \in D_\infty \mid k < i}$, that is the set of the pairs in $E_i$ that will be removed at stages later than $i$.
By Lemma \ref{lem:Eqi-inclusione} $\Set{(x,k) \in D_\infty \mid k < i} \subseteq E_i$, hence if $\Delta_i = \emptyset$ then $(\mbox{\rm Eq}_i)$ holds.
Suppose instead that $\Delta_i \neq \emptyset$ and  consider $(y,h) \in \Delta_i$ with minimal $h$.
$(y,h)$ is minimal in any $\Delta_j$ with $j\geq i$ to which it belongs, because the elements added to $\Delta_i$ have the form
$(z,p)$ for some $p>i$.

This implies that $h<i$ because $(y,h) \in E_i$, and
there exists $j>i$ such that $(y,h) \not\in D_j$ because $(y,h) \not \in D_\infty$. 

But if $D_j$ is the first such set, we have that $\Truth(z, X_j \cup \Set{x_j}) = \falseVal$ (where $x_j$ is the answer chosen at instruction 3. of step $j$) for some $(z,k) \in D_j$ such that $\Level(z) \leq \Level(y)$ and $k \leq h$. If we remove $(y,h)$, for the minimality of $h$ in $\Delta_j$ and the condition in the instruction 4. we remove all elements of $\Delta_j$. Therefore $\Delta_j = \emptyset$, and hence $(\mbox{\rm Eq}_j)$ holds. \qed

Let $X_{i_0}, X_{i_1}, \ldots$ be the subsequence of $X_0,X_1,\ldots$ such that $(\mbox{\rm Eq}_{i_k})$ holds for all $i_k$.
This is an infinite sequence by Lemma \ref{lem:inclusione_infOften}, and it is a non decreasing chain w.r.t. subset inclusion by construction.

\begin{lemma}\label{lem:r-empty}
~
\begin{enumerate}
\item \label{lem:r-empty-a} For some $k_0, z_1, \ldots, z_n$ and for all $k \geq k_0$  we have $r(X_{i_k}) = \Set{z_1,\ldots,z_n}$.
\item \label{lem:r-empty-b} For every $x\in\Atom$ either $\Truth(x,X_i) = \trueVal$ definitely or $\Truth(x,X_i) = \falseVal$ definitely.
\item \label{lem:r-empty-c} There are $k_1, \ldots , k_n \geq k_0$ such that $\Truth(z_i,X_i) = \trueVal$  for all $i \geq k_j$.
\item \label{lem:r-empty-d} For all $ k \geq k_j$ if $z_j \in X_{i_k + 1} \setminus X_{i_k}$ then $(z_j , i_k) \in D_{\infty}$.
\item \label{lem:r-empty-e} $r(X_{i_k}) = \emptyset$ for all $k \geq \max(k_1, \ldots, k_n)$.
\end{enumerate}
\end{lemma}

\proof Part (\ref{lem:r-empty-a}) follows by Lemma \ref{lem:stab_mon} and the fact that the subsequence $X_{i_0}, X_{i_1}, \ldots$ is a non decreasing chain. \\

\noindent
Part (\ref{lem:r-empty-b}) is a consequence of the fact that $\Truth(x,\_)$ is continuous in its second argument, and it is layered, so that it depends only
on the answers in the chain $X_0\restr 1 \subseteq X_1\restr 1 \subseteq \cdots $. \\

Part (\ref{lem:r-empty-c}): by part (\ref{lem:r-empty-a}) and Lemma \ref{lem:inclusione_infOften}, every $z_j$ belongs to infinitely many $r(X_i)$, since $X_{i_0}, X_{i_1}, \ldots$ is a subsequence of $X_0,X_1,\ldots$. Then,
by (\ref{lem:r-empty-b}), we have $\Truth(z_j,X_i) = \trueVal$ definitely. Indeed we cannot have $\Truth(z_j ,X_i)=\falseVal$ definitively since $r$ is a realizer, and therefore $\Truth(z_j ,X_{i_k}) = \trueVal$ for all $k$. \\

\noindent
Part (\ref{lem:r-empty-d}):
assume $k \geq k_j$ and $z_j \in X_{i_k+1} \setminus X_{i_k}$, in order to prove $(z_j, i_k) \in D_{\infty}$. To this aim, we assume $i \geq i_k + 1$ in order to prove $(z_j, i_k) \in D_i$. By definition of $D_i$, we have to prove that if $D_{i} \setminus D_{i-1} = {(x_{i-1},i-1)}$, then for all $(y,h) \in D_i$ with $h\leq i_k$ we have $\Truth(y, X_{i-1} \cup \Set{x_i}) = \trueVal$. If $h<i_k$, then, since $i \leq i_k + 1$, any $(y,h) \in D_i$ is in $X_{i_k} \subseteq D_{\infty}$, therefore $\Truth(y, X_{i-1} \cup \Set{x_i}) = \trueVal$. Assume $h = i_k$, hence $y = z_j$: we have to prove that $\Truth(z_j, X_{i-1} \cup {x_i}) = \trueVal$. From the fact that $\Truth$ is stratified we deduce $\Truth(z_j, X_{i-1} \cup {x_i}) = \Truth(z_j, X_{i-1} \cup {x_i} \restr 1) = \Truth(z_j,X_i \restr 1 ) = \Truth(z_j,X_i) = \trueVal$ by the choice of $z_j$.\\

\noindent
Part (\ref{lem:r-empty-e}): let $r(X_{i_k}) = \Set{z_1,\ldots,z_n}$ be the definitely constant value of $r$ over the chain $X_{i_0}, X_{i_1}, \ldots$, existing by
part (\ref{lem:r-empty-a}). By contradiction suppose that $n > 0$: by the choice of instruction $3.$ of the algorithm
there is some $z_j \in X_{i_k + 1}\setminus X_{i_k}$. By Part (\ref{lem:r-empty-d})
we have $(z_j, i_k) \in D_{\infty}$, therefore $z_j \in X_{i_h} \cap r(X_{i_h})$  for all $h$ large enough, contradicting the fact that $r$ is a realizer. \qed

Lemma \ref{lem:r-empty} (\ref{lem:r-empty-e}) and the fact that $X_{i_0}, X_{i_1}, \ldots$ is an infinite subsequence of $X_0,X_1,\ldots$ implies the termination of the algorithm. In summary we have:

\begin{theorem}\label{thr:algoSoundTerm}
If $r\in \Realizer$ is a computable realizer over a computable layered knowledge structure with knowledge space $\InfState$,
then for all
finite sound state $X\in\State$ the algorithm {\bf Find Sound Zero} always terminates and computes a sound and finite zero of $r$.
\end{theorem}

We observe that the use of the ``timing'' variable $n$ is essential to the proof of termination. Indeed it encodes, though in a very crude way, the logical dependency of hypothesis from other hypotheses: when the latter are falsified then the algorithm deletes all the subsequent hypothesis since they might depend on the falsified ones. This has the effect to forbid adding to the knowledge state infinitely many hypothesis, which is at the basis of the termination proof.

The algorithm and the proof we have presented suffer of some limitations. The less severe one is the limitation of $\Level$ to the ordinal $2$:
in \cite{Berardi-deLiguoro:FICS13} we have extended the construction to knowledge structures of any finite level. A further limitation is the (non deterministic) choice of a single answer from $r(X)$ at each step, which forces a sequential strategy instead of a parallel
one. A more efficient choice would be a consistent non-empty subset $U$ included in $r(X)$, whenever there is some.
Also the idea of using (linear) time to avoid infinite choices of the hypothesis is not the optimal solution to the problem of removing  the hypothesis that logically depend on the falsified ones. We leave the improvement of the algorithm to further work.

\section{Concluding remarks and further work}

We have defined the notions of state of knowledge and of state topology. We have then redefined in the more general setting of non-monotonic learning, the concepts of individual (here called  ``solution'') and of interactive realizer that we treated elsewhere, proving the completeness of learnability w.r.t. validity, which is the counterpart of classical truth in the present setting.

The definitions and results obtained are aimed at the development of a full theory of learning strategies and of their convergence properties, which is work in progress. We also observe that the solution and the realizer illustrated in the examples of \S \ref{sec:realizer} are crude simplifications of the learning strategy implicit in the example of \S \ref{sec:learning}, which is capable of using the counterexamples in a more ingenuous and efficient way. The investigation of the interpretation of classical proofs in terms of learning strategies is a natural further step, extending the work we have done in the monotonic case.

Since learning strategies working with finite approximations are effective  (and indeed we have shown that finite and sound knowledge states exist and suffice), a question of efficiency of the algorithms one extracts from proofs with our method is naturally there, together with the analysis of suitable data structures representing time and logical dependancies, which are essential to complete the present approach.

%\bibliographystyle{plain}% the recommended bibstyle
%\bibliography{LearnBib}

\end{document}